# Assessment of Diagnostic Capabilities of Methods of Recreation of Voltage Fluctuations

Piotr Kuwałek
*Institute of Electrical Engineering and Electronics*
*Poznań University of Technology*
Poznań, Poland
piotr.kuwalek@put.poznan.pl

*Abstract*— Voltage fluctuations are one of the most common low-frequency disturbances of power quality. Diagnostics of these disturbances are a complex process because voltage fluctuations affect different loads in different ways. Therefore, there is no measure of power quality that allows for the complementary assessment of severity of this disturbance, allow for the identification of sources of voltage fluctuations, and post-factum investigation of their effects. Among the currently used measures of voltage fluctuations, voltage fluctuation indices have the greatest diagnostic capabilities. Many preliminary studies also show the potential possibility of recreation of voltage fluctuations, including: based on voltage fluctuation indices. This paper presents the results of research on methods of recreation of voltage fluctuations from voltage fluctuation indices. The research carried out included a set of data obtained in a real power grid. Moreover, the impact of the discrimination period on the accuracy of recreation of voltage fluctuations has been assessed. The presented research results show, on the one hand, the usefulness of voltage fluctuation indices in the process of recreation of voltage fluctuations and, on the other hand, further challenges in the recreation of voltage fluctuations.

## I. INTRODUCTION

Voltage fluctuations are one of the most common low-frequency disturbances of power quality [1,2]. Voltage fluctuations affect different electrical loads in different ways [3-6], making the diagnostics of this disturbance a complex task [7-9]. The problem even occurs in measuring the severity of this disturbance [10-14]. Many measures of voltage fluctuations include: short-term flicker indicator $P_{st}$ [15], $\Delta V_{10}$ indicator [16], voltage fluctuation indices (VFI): the amplitude $\delta U$ and the rate $f$ of voltage fluctuations [17]. Among the mentioned measures of voltage fluctuations, VFI have the greatest diagnostic capabilities. VFI allow for the assessment of the effects of voltage fluctuations on different types of loads (not only light sources - flicker assessment) [17]. In recent years, it has also been noticed that the VFI with other currently recorded measures of power quality allows for the recreation of voltage fluctuations [18-21]. The issue of recreation of voltage fluctuations is very useful because it allows, for example:

- conversion of VFI into other measures of voltage fluctuations [20];
- post-factum investigation of different types of loads (analysis of potential effects of voltage fluctuations of a specific nature) [21];

This work was funded by National Science Centre, Poland – 2021/41/N/ST7/00397. For the purpose of Open Access, the author has applied a CC–BY public copyright licence to any Author Accepted Manuscript (AAM) version arising from this submission.

- implementation of complex algorithms in offline mode without the need to modify the existing measurement and recording equipment [22].

However, it is worth noting some limitations in the results currently presented regarding the recreation of voltage fluctuations from VFI [18-21]. So far, the recreation process has mainly involved laboratory tests in a controlled environment, where the voltage signal to be analyzed was obtained from an arbitrary generator. Therefore, so far, the precision of the recreation of voltage fluctuations has been mainly analyzed on the example of a certain idealized group of signals [18-21]. Additionally, when analyzing the research results available in the literature, it can be noticed that in the process of recreation of voltage fluctuations, only the aggregation of measures for a relatively long recording period has been considered. This article presents selected results of research in the field of assessment of the diagnostic capabilities of methods of recreation of voltage fluctuations from VFI, which fill the research gap. In this regard, a comparison is made in the errors of recreation of voltage fluctuations on a set of test signals obtained from the real low-voltage network [23]. The data set used in the investigation includes disturbance situations from a different number of dominant disturbing loads. Moreover, on the same set of test signals, the impact of the discrimination period of the VFI on the accuracy of recreation of voltage fluctuations is assessed. The results of the presented research show, on the one hand, the usefulness of VFI in the recreation of voltage fluctuations, and, on the other hand, they indicate further challenges in the recreation of voltage fluctuations.

## II. DESCRIPTION OF VOLTAGE FLUCTUATION INDICES

VFI are a measure that describes voltage fluctuations using the amplitude $\delta U$ and the rate $f$ of voltage fluctuations [17]. VFI are determined from changes in the rms values of voltage $U(t)$. Analogous VFI based on the voltage envelope $u_{env}(t)$: $\delta U_e$ and $f_e$, determined from the changes in the estimated voltage envelope $u_{env}(t)$ by demodulator with estimation of the signal carrier [24], have much better diagnostic properties [18]. However, such an approach based on the voltage envelope requires significant changes in the signal chains of the current-used measuring and recording devices [18]. Therefore, for the purposes of this paper, the conventional method for determining VFI is used [21]. The conventional approach creates the direct possibility of offline application of complex algorithms on recreated voltage fluctuations without modifying the currently used measurement and recording devices [21,22]. VFI: $\delta U$ and $f$, directly describe voltage changes qualified as voltage fluctuations. VFI measurement can be divided into two stages:

- detection of the amplitudes $\delta V$ of subsequent changes in the rms values of the voltage $U(t)$ in a given discrimination period $T_w$,

- statistical evaluation of the amplitudes $\delta V$ of subsequent changes detected during the period $T_w$, allowing for the determination of final form of VFI: the amplitude $\delta U$ and the rate $f$ of voltage fluctuations.

Detection of the amplitudes $\delta V$ of subsequent changes in the rms values of voltage $U(t)$ requires a preliminary calculation of the speed $SR$ of changes in the rms values of voltage $U(t)$. The change in the rms values of voltage $U(t)$ is classified as an amplitude $\delta V$ only when its speed $SR$ exceeds the limit value, which is usually taken in measuring and recording instruments to be equal to $1\%U_N$/s, where $U_N$ is the nominal rms value in the power grid. Based on the set of amplitude $\delta V$ values of subsequent changes in the rms values of voltage $U(t)$, statistics are performed that count the number of occurrences of individual $\delta V$ amplitudes in the subsets related to the maximum $\delta V$ value in a given period $T_w$, which is assumed as $\delta U$. For the purposes of this paper, the following subranges $\delta U$: [1.0,0.9], (0.9,0.8], (0.8,0.7], (0.7,0.5], (0.5,0.3], (0.3,0.1], (0.1,0.0), which for simplicity is written as: $f_{1.0-0.9}$, $f_{0.9-0.8}$, $f_{0.8-0.7}$, $f_{0.7-0.5}$, $f_{0.5-0.3}$, $f_{0.3-0.1}$, $f_{0.1-0.0}$. An example of determination of amplitudes $\delta V$ of subsequent changes in the rms values of voltage $U(t)$ is shown in Fig. 1.

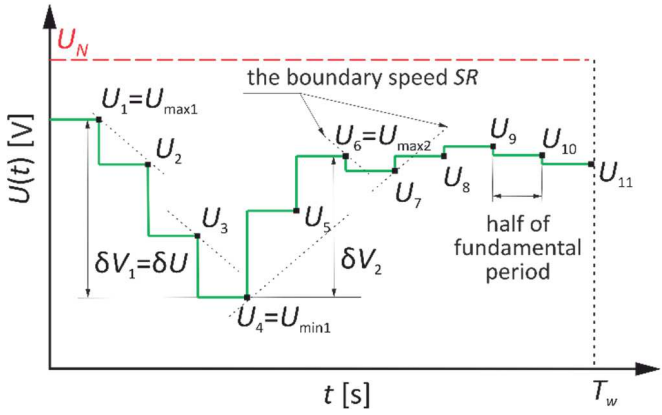

Fig. 1. The example of determination of amplitudes $\delta V$ of subsequent changes in the rms values of voltage $U(t)$

### III. Selected Methods of Recreation of Voltage Fluctuations

Methods of recreation of voltage fluctuations marked as M1, M2, M3 [18,21] are used for the purposes of research. Input data for individual methods (for period of discrimination $T_w$ equal to 1 second, 10 seconds, 30 seconds, 1 minute, 2 minutes, 5 minutes, 10 minutes) are: the minimum rms value of voltage $U_{min}$ in $T_w$; the maximum rms value of voltage $U_{max}$ in $T_w$; the average rms value of voltage $U_{avg}$ in $T_w$; the amplitude of voltage fluctuations $\delta U$; the rate of voltage fluctuations for individual subranges $\delta U$: $f_{1.0-0.9}$, $f_{0.9-0.8}$, $f_{0.8-0.7}$, $f_{0.7-0.5}$, $f_{0.5-0.3}$, $f_{0.3-0.1}$, $f_{0.1-0.0}$. In the recreation process, for research purposes, the following amplitudes $\delta V$ of particular changes in the rms values of voltage $U(t)$ are assumed to be equal to:

- $\delta U$ in the subrange $f_{1.0-0.9}$;
- $0.85\delta U$ in the $f_{0.9-0.8}$ subrange;
- $0.75\delta U$ in the $f_{0.8-0.7}$ subrange;
- $0.69\delta U/0.6\delta U/0.5\delta U$ in the subrange $f_{0.7-0.5}$ (the number of changes in the subrange divided into 3 equal sets, the rest falls on the central value);
- $0.49\delta U/0.4\delta U/0.3\delta U$ in the subrange $f_{0.5-0.3}$ (the number of changes in the subrange divided into 3 equal sets, the rest falls on the central value);
- $0.29\delta U/0.2\delta U/0.1\delta U$ in the subrange $f_{0.3-0.1}$ (the number of changes in the subrange divided into 3 equal sets, the rest falls on the central value);
- $0.09\delta U/0.05\delta U/0.01\delta U$ in the subrange $f_{0.1-0.0}$ (the number of changes in the subrange divided into 3 equal sets, the rest falls on the central value).

Generally, each method recreates subsequent changes in the rms values of voltage $U(t)$ in such a way that they oscillate around the average rms value of voltage $U_{avg}$ and so that subsequent changes do not leave the range: $U_{min}$ and $U_{max}$. Detailed differences in the process of recreation of voltage fluctuations by individual methods M1-M3 can be presented as follows [21].

**M1)** This method recreates voltage fluctuations as step changes in the rms values of voltage $U(t)$, which corresponds to the amplitude modulation with a square wave. The individual adopted amplitudes $\delta V$ of voltage changes are distributed evenly in the considered discrimination period $T_w$, and the order of their occurrence is random according to a uniform distribution. The subsequent assumed amplitudes $\delta V$ of voltage changes are distributed in such a way (by considering the sign of the voltage change) that they oscillate around the measured average rms value $U_{avg}$ and that the recreated values within the range defined by the measured minimum $U_{min}$ and maximum $U_{max}$ rms values of voltage.

**M2)** This method recreates voltage fluctuations as trapezoidal changes in the rms values of the voltage $U(t)$ assuming a constant speed of voltage fluctuation equal to $SR=300\%\delta U_N$/s. The individual adopted amplitudes $\delta V$ of voltage changes are evenly distributed in the considered discrimination time $T_w$ considered, and the order of their occurrence is random according to a uniform distribution. The subsequent adopted amplitudes $\delta V$ of voltage changes in the rms values $U(t)$ are distributed in such a way (by considering the sign of the voltage change) that they oscillate around the measured average rms value $U_{avg}$ and that the values of the recreated values within the range defined by the measured minimum $U_{min}$ and maximum $U_{max}$ rms values of voltage.

**M3)** This method recreates voltage fluctuations as trapezoidal changes in the rms values of voltage $U(t)$ assuming a variable speed $SR$ for recreation of voltage fluctuations. Individual speed $SR$ for subsequent introduced $\delta V$ amplitudes of voltage changes are randomized according to the gamma distribution with a shape parameter of 1 and a scale parameter of 300. The individual adopted amplitudes $\delta V$ of voltage changes are evenly distributed in the considered discrimination time $T_w$ considered, and the order of their occurrence is random according to a uniform distribution. The subsequent assumed amplitudes $\delta V$ of voltage changes are distributed in such a way (considering the sign of the change) that they oscillate around the measured average rms value $U_{avg}$ and that the recreated values within the range defined by the measured minimum $U_{min}$ and maximum $U_{max}$ rms values of voltage.

## IV. RESEARCH RESULTS AND DISCUSSION

For research purposes, a set of test signals [23] was used, which contains recorded sampled voltage values with a sampling rate of 20 kSa/s for a period of 10 minutes. The individual voltage signals in the test set were acquired in a real low-voltage power grid with a branching radial topology. A diagram of the power grid with its individual measurement points where the test signal acquisition process was carried out is shown in Fig. 2.

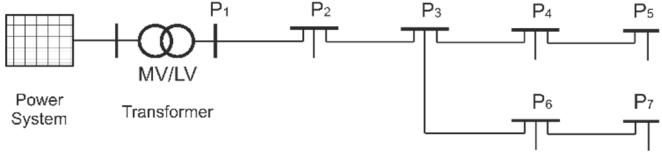

Fig. 2. The diagram of the power grid with its individual measurement points where the test signal acquisition process was carried out

The following parameters were determined for each voltage signal:

- average $U_{avg}$, minimum $U_{min}$, and maximum $U_{max}$ rms values of voltage;
- VFI: $\delta U$ and $f$;
- short-term flicker indicator $P_{st}$.

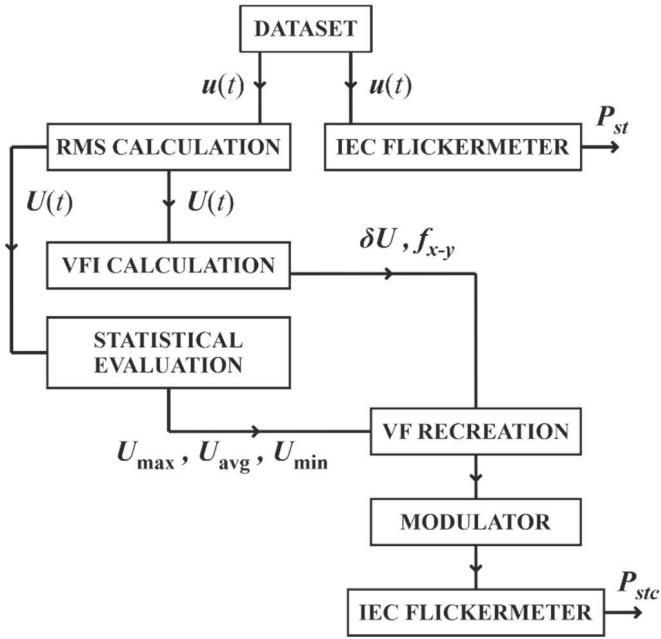

Fig. 3. The diagram showing the individual stages of signal processing

Individual parameters (except the indicator $P_{st}$) were determined with the discrimination period of 1 second, 10 seconds, 30 seconds, 1 minute, 2 minutes, 5 minutes, 10 minutes. The indicator $P_{st}$ was determined for 10 minutes. Taking into account the fact that coding information about voltage fluctuations in the form of VFI is a lossy process, it was assumed that the assessment of the accuracy of the individual investigated methods M1-M3 is reduced to the assessment of the possibility of recreation of such voltage fluctuations, the effects of which are analogous to those that occurred originally. Therefore, the indicator $P_{st}$ is used as a reference value that allows assessing the effects of voltage fluctuations in the form of flicker [18,21]. For the individual cases considered, the characteristics $P_{stc}=\mathbf{f}(P_{st})$ are determined, where the $P_{stc}$ value is the determined value of the indicator $P_{st}$ for the recreated voltage fluctuation signal. In the ideal case, $P_{sct}$ should be equal to $P_{st}$, resulting in the characteristic $P_{stc}=\mathbf{f}(P_{st})$ being linear with a slope coefficient $a_{Pst}$ [25] equal to 1 [20]. The greater the deviation of the coefficient $a_{Pst}$ from the value equal to 1, the greater the error. Additionally, the r-Pearson coefficient $r_{Pst}$ [26] was determined, defining the linear correlation between $P_{stc}$ and $P_{st}$, to exclude from the inference the situation in which the coefficient $a_{Pst}$ has a value close to 1, but there is a large dispersion of results between $P_{stc}$ and $P_{st}$ (coefficient $r_{Pst}$ has a value tending to zero). The closer the value of the coefficient $r_{Pst}$ is to 1, the greater the convergence of the $P_{stc}$ and $P_{st}$ results. A diagram showing the individual stages of signal processing [20] is shown in Fig. 3.

TABLE I. LIST OF CALCULATED COEFFICIENTS FOR EACH TEST SIGNALS

| $T_w$ [s] | $r_{Pst}$ [-] | | | $a_{Pst}$ [-] | | |
|---|---|---|---|---|---|---|
| | M1 | M2 | M3 | M1 | M2 | M3 |
| 1 | 0.898 | 0.892 | 0.957 | 0.889 | 0.791 | 0.967 |
| 10 | 0.905 | 0.897 | 0.953 | 0.894 | 0.774 | 0.980 |
| 30 | 0.906 | 0.899 | 0.953 | 0.899 | 0.777 | 0.986 |
| 60 | 0.905 | 0.899 | 0.952 | 0.901 | 0.777 | 0.990 |
| 300 | 0.898 | 0.892 | 0.951 | 0.901 | 0.775 | 1.002 |
| 600 | 0.895 | 0.891 | 0.950 | 0.901 | 0.774 | 1.007 |

Table I shows the results of the calculated coefficients $r_{Pst}$ and $a_{Pst}$ for the individual discrimination periods considered. On the basis of the results obtained, it can be seen that regardless of the adopted discrimination period, the dispersion between the calculated coefficients $r_{Pst}$ is small, and the dispersion between the calculated coefficients $a_{Pst}$ is also small. Therefore, it can be indicated that regardless of the discrimination period, individual methods make it possible to recreate voltage fluctuations whose flicker is highly convergent (significant correlation coefficient) with the flicker caused by the original voltage fluctuations. For the voltage fluctuations of varying severity considered, it can be seen that the method M3, based on trapezoidal voltage fluctuations, the speed $SR$ of which is randomized according to the gamma distribution, is most effective. The effectiveness of this approach results from the disturbance random nature of the sources that occur in the real power grid. The methods M1 and M2 result in an underestimation of the $P_{st}$ measurement result, so the recreated voltage fluctuations cause a smaller flicker. The worst efficiency occurs for the method M2, where a finite constant speed $SR$ of voltage changes is assumed. The error in this method is related to the adoption of too low a value for the speed of voltage change, because slow voltage changes result in less severe effects of voltage fluctuations [21,22].

Table II and Table III present the results of the calculated coefficients $r_{Pst}$ and $a_{Pst}$ for the individual discrimination periods and for test signals whose $P_{st}$ values are smaller and larger than 2, respectively. This division aims to better analyze the effectiveness of the considered methods, as it allows for the assessment of the performance of individual methods at the limit of normative requirements (in the low-voltage network, according to the standard EN 50160 [27], the limit $P_{st}$ value is equal to 1) and in the event of high voltage fluctuations. In both cases, there is still a negligible impact of discrimination time on the effectiveness of individual methods. Furthermore, the method M3 still has the best results. However, it is worth noting that for limit states (with relatively low severity of voltage fluctuations), the effectiveness of the methods M1 and M2 increases significantly.

TABLE II. LIST OF CALCULATED COEFFICIENTS FOR TEST SIGNALS FOR WHICH SHORT-TERM FLICKER INDICATOR IS LESS THAN 2

| $T_w$ [s] | $r_{Pst}$ [-] | | | $a_{Pst}$ [-] | | |
|---|---|---|---|---|---|---|
| | M1 | M2 | M3 | M1 | M2 | M3 |
| 1 | 0.946 | 0.945 | 0.912 | 0.987 | 0.943 | 0.998 |
| 10 | 0.952 | 0.954 | 0.919 | 0.970 | 0.914 | 1.000 |
| 30 | 0.951 | 0.954 | 0.916 | 0.973 | 0.916 | 1.010 |
| 60 | 0.951 | 0.953 | 0.914 | 0.974 | 0.917 | 1.019 |
| 300 | 0.948 | 0.951 | 0.908 | 0.978 | 0.920 | 1.058 |
| 600 | 0.947 | 0.949 | 0.899 | 0.980 | 0.922 | 1.082 |

TABLE III. LIST OF CALCULATED COEFFICIENTS FOR TEST SIGNALS FOR WHICH SHORT-TERM FLICKER INDICATOR IS GREATER THAN 2

| $T_w$ [s] | $r_{Pst}$ [-] | | | $a_{Pst}$ [-] | | |
|---|---|---|---|---|---|---|
| | M1 | M2 | M3 | M1 | M2 | M3 |
| 1 | 0.833 | 0.823 | 0.928 | 0.887 | 0.788 | 0.966 |
| 10 | 0.841 | 0.829 | 0.922 | 0.892 | 0.772 | 0.980 |
| 30 | 0.843 | 0.831 | 0.922 | 0.897 | 0.774 | 0.986 |
| 60 | 0.842 | 0.831 | 0.921 | 0.899 | 0.774 | 0.989 |
| 300 | 0.830 | 0.821 | 0.919 | 0.899 | 0.772 | 1.001 |
| 600 | 0.826 | 0.819 | 0.918 | 0.899 | 0.771 | 1.005 |

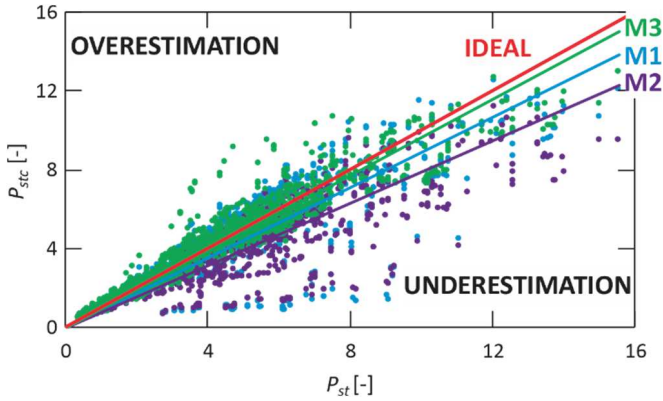

Fig. 4. Research results for each test signal with a discrimination period $T_w$ of 1 second

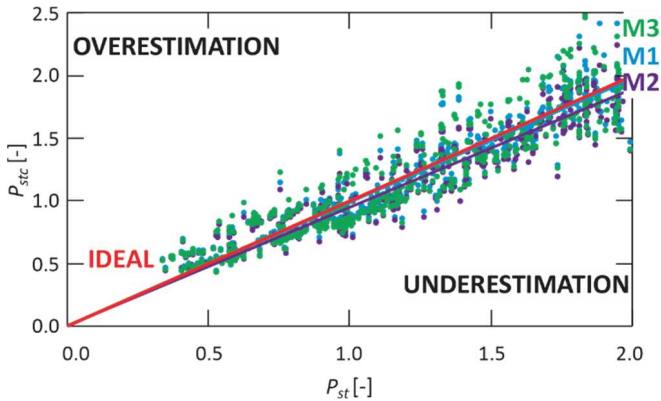

Fig. 5. Research results for test signals with a discrimination period $T_w$ of 1 second for which the $P_{st}$ values are less than 2

Figs. 4-6 present examples of the research results for the discrimination period $T_w$ in the form of characteristics $P_{stc}=\mathbf{f}(P_{st})$. The scatter of the results obtained for individual test signals can be observed in the individual characteristics. For other discrimination periods $T_w$, the scatter is analogous; therefore, considering the limitation in the number of pages, the presentation of characteristics $P_{stc}=\mathbf{f}(P_{st})$ for the remaining considered discrimination periods $T_w$ is omitted. The discrepancies can be related to the fact that, in the process of a recreation of voltage signal with voltage fluctuations, the carrier signal is assumed to be sinusoidal, while in reality the signal is usually distorted. In recent years, the simultaneous occurrence of voltage fluctuations and voltage distortions has been shown to affect the value of the indicator $P_{st}$ and consequently the flicker [28].

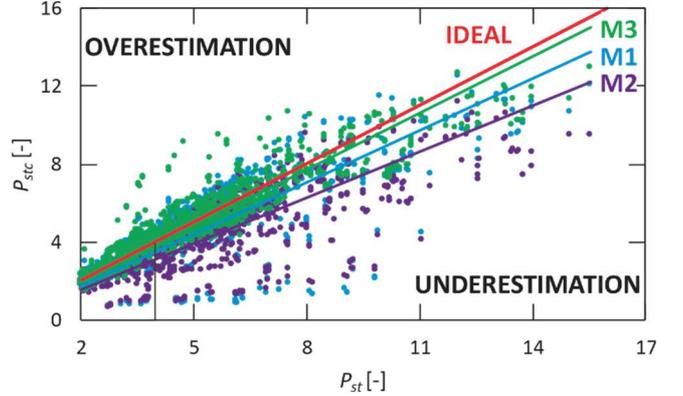

Fig. 6. Research results for test signals with a discrimination period $T_w$ of 1 second for which the $P_{st}$ values are greater than 2

V. CONCLUSION

The paper presents the results of the research on methods of recreation of voltage fluctuations from VFI. The research carried out included a set of data obtained in a real power grid. The research evaluated the effectiveness of the recreation of voltage fluctuations by the individual methods tested. The effectiveness of recreation of voltage fluctuations is assessed by the ability to recreation of the effects of voltage fluctuations in the form of a flicker analogous to the effects caused by the original voltage fluctuations. The indicator $P_{st}$ is used to assess diagnostic possibilities. The research also assesses the impact of the duration of discrimination on the accuracy of recreation of voltage fluctuations. The presented research results show that regardless of the adopted discrimination period, the dispersion between the calculated coefficients $r_{Pst}$ is small, and the dispersion between the calculated coefficients $a_{Pst}$ is also small. Therefore, it can be indicated that regardless of the discrimination period, individual methods make it possible to recreate voltage fluctuations whose flicker is highly convergent (significant correlation coefficient) with the flicker caused by the original voltage fluctuations. Among the methods tested, the most effective is the method M3, which is based on trapezoidal voltage fluctuations, the speed $SR$ of which is randomized according to the gamma distribution. The effectiveness of this approach results from the disturbance random nature of the sources that occur in the real power grid. However, it is worth noting that for limit states (with relatively low severity of voltage fluctuations), the effectiveness of the methods M1 and M2 increases significantly. The presented research results show, on the one hand, the usefulness of VFI for the recreation of voltage fluctuations, and, on the other hand, further challenges in the recreation of voltage fluctuations.